\documentstyle [12pt] {article}
\begin{document}


\pagestyle{empty}
\rightline{\vbox{
\halign{&#\hfil\cr
&hep-ph/9407341\cr
&UCD-94-26\cr
&June 1994\cr}}}
\bigskip
\bigskip
\bigskip

{\bf \centerline{Helicity Probabilities For Heavy Quark Fragmentation Into
Excited Mesons}}

\bigskip
\normalsize

\centerline{Tzu Chiang Yuan}
\centerline{\sl Davis Institute for High Energy Physics}
\centerline{\sl Department of Physics, University of California,
    Davis, CA  95616}
\bigskip

\begin{abstract}

In the fragmentation of a heavy quark into a heavy meson whose
light degrees of freedom have angular momentum $3/2$, all the helicity
probabilities are completely determined in the heavy quark limit up
to a single probability $w_{3/2}$. We point out that this probability
depends on the longitudinal momentum fraction $z$ of the meson and on its
transverse momentum $p_\bot$ relative to the jet axis.
We calculate $w_{3/2}$ as a function of scaling variables corresponding
to $z$ and $p_\bot$ for the heavy quark limit of the perturbative
QCD fragmentation functions for $b$ quark to fragment into $(b \bar c)$ mesons.
In this model, the light degrees of freedom prefer to have
their angular momentum aligned transverse to, rather than along, the jet axis.
Implications for the production of excited heavy mesons,
like $D^{**}$ and $B^{**}$, are discussed.

\end{abstract}

\vfill\eject\pagestyle{plain}\setcounter{page}{1}

The discovery of the powerful heavy quark spin-flavor
symmetry \cite{wiseguys} in Quantum Chromodynamics (QCD) and the development
of the Heavy Quark Effective Theory (HQET) \cite{hqet}
have greatly improved our theoretical understanding of
hadrons containing a single heavy quark $Q$.
The crucial idea in HQET is that both the heavy quark mass and
spin decouple from the strong interaction dynamics between the heavy quark
and the light degrees of freedom in the limit of infinitely heavy quark mass.
The decouplings of the mass and spin occur because
in the limit $m_Q \to \infty$ the only variable to describe a free
propagating heavy quark is its 4-velocity $v$. In the presence of
gauge fields, the propagation of the heavy quark is simply described by the
Wilson line $P \exp i \int A_{\mu} v^{\mu} dt$,
which contains no information about the heavy quark flavor nor its spin.
Furthermore, the leading operator -- the chromo-magnetic dipole
moment that couples heavy quark spin to the gluon field is inversely
proportional to the heavy quark mass. To the first approximation
in the heavy quark mass expansion, a heavy quark can be treated as a
static color source for the remaining light degrees of freedom that
make up the physical heavy-light hadrons observed in Nature.

Falk and Peskin \cite{falkpeskin} have recently pointed out that
heavy quark spin symmetry provides strong constraints on the helicity
probabilities of the heavy mesons produced by the
fragmentation/hadronization of a heavy quark.
For heavy mesons whose light degrees of freedom have angular momentum
1/2, the helicity probabilities are completely determined in the heavy quark
limit. If the light degrees of freedom have angular momentum 3/2, the
probabilities are determined up to a single parameter $w_{3/2}$.
In this paper, we point out that the probability $w_{3/2}$ depends on the
longitudinal momentum fraction $z$ of the heavy meson relative to the heavy
quark jet and its transverse momentum $p_\bot$ relative to the jet axis.
We calculate $w_{3/2}$ as a function of the scaling variables
corresponding to $z$ and $p_\bot$ for the $m_b \to \infty$ limit
of the perturbative QCD (PQCD) fragmentation functions for a
$b$ quark to fragment into $(b \bar c)$ mesons. These fragmentation functions
can be used as a model for the fragmentation of a heavy quark into heavy-light
mesons. The implications for the production of excited heavy-light mesons,
like the $D^{**}$ and $B^{**}$, are discussed.

In the heavy quark limit it is convenient
to label  hadronic  states
by the eigenvalues $j$ and $j_l$ of the total angular momentum
$\vec J$ of the hadron
and the angular momentum $\vec J_l$ of the light degrees of freedom
respectively \cite{spectrum}. In general, the spectrum
of hadrons containing a single heavy quark $Q$ has,
for each $j_l$, a degenerate doublet $(j_-,j_+)$ with
total angular momentum $j_\pm = j_l \pm 1/2$.
(For the exceptional case of a baryon with $j_l=0$, there is only a
singlet with total angular momentum $1/2$.)
For the heavy-light $(Q \bar q)$ mesons,
one can write $\vec J_l = \vec S_q + \vec L$
where $\vec S_q$ is the spin of the valence light antiquark $\bar q$
and $\vec L$ is the orbital angular momentum with integer eigenvalue $L$.
For $S$-wave heavy-light mesons,
$L=0$ and ${j_l}^P={1 \over 2}^-$, we have the familiar degenerate doublet
$({j_-}^P,{j_+}^P)=(0^-,1^-)$ consisting of a pseudoscalar and a vector meson.
For $P$-wave heavy-light mesons,
$L=1$ and the ${j_l}^P$ of the light degrees of freedom can be either
${1 \over 2}^+$ or ${3 \over 2}^+$.
In this case, we have two distinct doublets
$({j_-}^P,{j_+}^P)\; = \; (0^+,1^{+'})$
and $(1^+,2^+)$ for ${j_l}^P$ = ${1 \over 2}^+$ and ${3 \over 2}^+$
respectively.
While the $0^+$ and $2^+$ states are identified as the $^3P_0$ and $^3P_2$
states constructed in the $LS$ coupling scheme respectively,
the $1^{+'}$ and $1^+$ states are linear combinations of
the $^1P_1$ and $^3P_1$ states:
$|1^{+'} \rangle  = \sqrt{1/3} |^1P_1 \rangle +  \sqrt{2/3} |^3P_1 \rangle$
and
$|1^+ \rangle   = - \sqrt{2/3} |^1P_1 \rangle + \sqrt{1/3} |^3P_1 \rangle$ .

Falk and Peskin \cite{falkpeskin} showed that
heavy quark spin symmetry combined with the parity invariance of QCD
interactions can impose very useful relations among the
probabilities for a heavy quark with a given
helicity to fragment/hadronize into the various helicity states of the
heavy-light hadrons within the same doublet.
For definiteness, we assume the heavy quark $Q$ is purely left-handed in
what follows.
For the case of $j_l = {1 \over 2}$, parity invariance implies that
the two fragmentation probabilities for the heavy quark $Q$
to hadronize by combining with light degrees of freedom
with helicity $m_l$ = $-1/2$ or +1/2 must be the same.
Although heavy quark symmetry allows coherent
superposition of the two zero helicities states in the
corresponding $(0,1)$ doublet at the very early stage of the fragmentation
process, the relative population of the four helicities states in the doublet
are completely determined by symmetry.
The resulting table of fragmentation probabilities is \cite{falkpeskin}
\begin{equation}
\label{p01}
\left(
\begin{array}{cc}
P_1(h) \\  \\ P_0(h)
\end{array}
\right)
=
\left(
\begin{array}{ccc}
{1 \over 2} & {1 \over 4} &  0 \\
 & & \\
 & {1 \over 4} &
\end{array}
\right)  \; \; ,
\end{equation}
where the helicity $h$ runs through the values $-1$, 0, +1 across the table.
However, for $j_l = {3 \over 2}$, parity invariance implies the following
table of fragmentation probabilities for a heavy quark
to fragment into various helicity states of the light degrees
of freedom \cite{falkpeskin}
\begin{equation}
\label{p32}
P_{3/2} ( m_l ) =
\left( {1 \over 2}w_{3/2},{1 \over 2}(1 - w_{3/2}),
 {1 \over 2}(1 - w_{3/2}),{1 \over 2}w_{3/2} \right) \; \; ,
\end{equation}
where the helicity $m_l$ of the light degrees of freedom
runs through the values $-{3 \over 2}$, $-{1 \over 2}$,
$+{1 \over 2}$, $+{3 \over 2}$ across the table.
The Falk-Peskin parameter, $w_{3/2}$, is
the conditional probability for the heavy quark to fragment into a
meson system whose light degrees of freedom are in the
maximum helicity states $m_l = \pm 3/2$.
It can take values between 0 and 1.
At the very  early stage in the fragmentation process,
heavy quark symmetry allows coherent linear superposition of the
various components in the $(1,2)$ doublet which have
the same helicity $h = m_l - {1 \over 2}$. However, at a later time
determined by the mass splittings within the doublet, the various helicity
components of the $(1,2)$ doublet will propagate incoherently,
resulting in the following table of
fragmentation probabilities \cite{falkpeskin}
\begin{equation}
\label{pdoublet}
\left(
\begin{array}{cc}
P_2(h) \\  \\ P_1(h)
\end{array}
\right)
=
\left(
\begin{array}{ccccc}
{1 \over 2} w_{3/2} & {3 \over 8}(1 - w_{3/2}) &
{1 \over 4} (1 - w_{3/2}) & {1 \over 8} w_{3/2} & 0 \\
 & & & & \\
 & {1 \over 8}(1 - w_{3/2}) & {1 \over 4}(1 - w_{3/2}) &
  {3 \over 8}w_{3/2} &
\end{array}
\right)  \; \; ,
\end{equation}
with the helicity $h$ runs through the values $-2$, $-1$, 0, +1, +2
across the table.
Therefore, all the relative fragmentation probabilities for the 8 helicity
states of a $(1,2)$ doublet are determined, up to the Falk-Peskin parameter
$w_{3/2}$. This non-perturbative parameter can not be
determined without a dynamical calculation.

Consider the leading order Feynman diagram shown in Fig.1 for
a heavy $b$ quark to fragment into a heavy-heavy $(b \bar c)$ meson
by creating a $c \bar c$ pair from the vacuum. In the limit of
$m_b/m_c \to \infty$, this can be taken
as a model for heavy quark fragmentation into heavy-light mesons.
Following our previous works in \cite{swave,bcfrags,bcfragp}
(see also \cite{chen}), the fragmentation function $D(z)$ can be expressed as
$D(z) = \int_{s_0(z)}^\infty ds \, D(z,s)$, with
$s_0(z) = M^2/z + r^2M^2/(1-z)$ and
$D(z,s) = (16 \pi^2)^{-1} \lim_{q_0/m_b \to \infty}
( |{\cal M}|^2 / |{\cal M}_0|^2 )$.
$\cal M$ is the amplitude for a high energy source
(symbolically denoted by $\Gamma$ in Fig.1) to create
$b^*(q) \to (b \bar c)(p) + c(p')$ with total 4-momentum $q = p + p'$,
and ${\cal M}_0$ denotes the same source creating an on-shell
heavy $b$ quark with the same 3-momentum $\vec q$.
In the nonrelativistic approximation, $M = m_b + m_c$; and we define
$r = m_c/M$. $s=q^2$ is the virtuality of the fragmenting $b$ quark.
In a frame where the virtual $b^*$ quark has 4-momentum
$q = (q_0,0,0,q_3)$, the longitudinal momentum fraction of $(b \bar c)$
is $z=(p_0+p_3)/(q_0+q_3)$ and its transverse momentum is
$\vec p_\bot = (p_1,p_2)$.
For the $b$ quark propagator and $bbg$ vertex that entered in Fig.1,
we will apply the techniques of HQET \cite{hqet}.
It is straightforward to write down the
following matrix elements ${\cal M}(^1P_1,h)$ and ${\cal M}(^3P_1,h)$
for $b \to (b \bar c)$, with the $(b \bar c)$ bound state
in definite helicity $h$ of the $^1P_1$ and $^3P_1$ states
respectively:
\begin{equation}
\label{M1p1}
{\cal M}(^1P_1,h) = i \delta_{ij}g^2C_F R'(0) \sqrt{3 \over 16 r^2 \pi N_c M}
{1 \over k^4 \, v \cdot k} \epsilon^*_\alpha(h) \bar u(p')
\gamma_5 V^{\alpha} \left( {1 \, + \not\! v \over 2} \right) \Gamma \; ,
\end{equation}
with the vertex $V^{\alpha}$ given by
\begin{equation}
\label{vertexv}
V^{\alpha} = 4 r M q^{\alpha}
\Biggl( 1 - {v \cdot k \over n \cdot k} \not\! n \Biggr)
- 2 r M {k^2 v \cdot k \over (n \cdot k)^2 } n^{\alpha} \not\! n
+ k^2 \Biggl( 1 +{v \cdot k \over n \cdot k} \not\! n \Biggr)
\gamma^{\alpha} \; ;
\end{equation}
and
\begin{equation}
\label{M3p1}
{\cal M}(^3P_1,h) = - \delta_{ij}
g^2 C_F R'(0) \sqrt{3  \over 32 r^2 \pi N_c M}
{1 \over k^4 \, v \cdot k} \epsilon_{\alpha\beta\mu\nu} v^\mu
\epsilon^{\nu *}(h) \bar u(p')
V^{\alpha\beta} \left( {1 \, + \not\! v \over 2} \right) \Gamma \; ,
\end{equation}
with the vertex tensor $V^{\alpha\beta} = \tilde V^\alpha \gamma^\beta$,
where $\tilde V^\alpha$ can be obtained from $V^\alpha$ in (\ref{vertexv})
with $\not\! n \to - \not\! n$ and $\gamma^\alpha \to - \gamma^\alpha$.
In (\ref{M1p1}) and (\ref{M3p1}),
$i$ and $j$ are the fundamental color indices, $N_c$ is the number of color,
$C_F = (N_c^2-1)/(2N_c)$, and
$R'(0)$ and $\epsilon (h)$ are the derivative of the radial wave function and
the helicity wave function for the bound state respectively.
We have picked the axial gauge associated with the vector
$n^\mu = (1, - \vec p/|\vec p|)$.
In this gauge, the short-distance process that creates the energetic
heavy quark $b^*$ (symbolically denoted by $\Gamma$ in Fig.1) and the
subsequent fragmentation of the heavy quark become manifestly
factorized.
In the above equations, we have used $p=M v$ and $q=m_b v + k$, where
$v$ and $k$ are the 4-velocity and residual momentum of the $b$ quark
respectively.  Some useful kinematical relations are
$k^2 = 2 r v \cdot k = r (s -  (1 - r)^2 M^2)/M$.
The tree-level matrix element squared $|{\cal M}_0|^2$ is given by
$N_c (M / z) {\rm Tr} [\Gamma \bar \Gamma ( 1 \, + \not\! v) ]$.
In writing down the above matrix elements, we have assumed
the $P$-wave bound state is a color-singlet. It has been pointed out recently
by Bodwin-Braaten-Lepage \cite{bbl} that,
beyond leading orders in the calculations of the production and decay
rates of $P$-wave quarkonium \cite{bbl,gfragp}
and $(b \bar c)$ mesons \cite{bcfragp},
there is also a color-octet $S$-wave mechanism needed to be
taken into account in order to avoid infrared divergencies
that spoil factorization. We will also include
the color-octet $S$-wave contributions in what follows.

Heavy quark fragmentation functions have nontrivial limits as
$m_Q \to \infty$ (or $r \to 0$) if they are expressed in terms of the
scaling variable
$y=(1/z - 1 + r)/r$, first used by Jaffe and Randall \cite{jaffe},
and the rescaled transverse momentum $t = |\vec p_\bot|/(r M)$.
We therefore define the fragmentation functions
$D(y,t)$, $D(t)$, and $D(y)$ according to  the following
changes of variables:
\begin{eqnarray}
\int_0^1 dz \, D(z) & = &
 \int_0^1 dz \int_{s_0(z)}^\infty ds \, D(z,s)  \; , \\
& = & \int_1^\infty dy \int_0^\infty dt \, D(y,t) \; , \\
& = & \int_0^\infty dt \, D(t) \; , \\
& = & \int_1^\infty dy \, D(y) \; .
\end{eqnarray}
The relation among $s$, $t$, and $z$ is given by
$s = M^2 [ (1 + r^2 t^2)/z + r^2(1+t^2)/(1-z) ]$.
Following the same procedures as in Refs.\cite{swave,bcfrags,bcfragp},
we square the matrix elements ${\cal M}(^1P_1,h)$
and ${\cal M}(^3P_1,h)$, calculate the interference between these
two amplitudes, and project out the transversely and longitudinally
polarized $1^+$ states, one can deduce the generalized
polarized fragmentation functions
$D_{T}(y,t) = D_T^{(1)}(y,t) + 2 \, D^{(8)}(y,t)$ and
$D_L(y,t) = D_L^{(1)}(y,t) + D^{(8)}(y,t)$, depending on both
variables $y$ and $t$.
In the limit $r \to 0$, the color-singlet pieces $D_{T,L}^{(1)}(y,t)$
are given by
\begin{eqnarray}
\label{dtyt}
D^{(1)}_T (y,t) &=& {4 C_1 \over 3} {(y-1)^2 t \over y^4 (t^2 + y^2)^6}
\left\{ (y-4)^2 y^6 + y^4(40-32y+26y^2-4y^3+y^4)t^2
\right. \nonumber \\
&& \;\; \left.
+ \; 3y^3(16-8y+2y^2+3y^3)t^4+ 3y^2(8-4y+3y^2)t^6 +(1+y)^2 t^8 \right\} \, ,
\end{eqnarray}
\begin{eqnarray}
\label{dlyt}
D^{(1)}_L (y,t) &=& {8 C_1 \over 3} {(y-1)^2 t \over y^4 (t^2 + y^2)^6}
\left\{ (y-4)^2 y^6 + y^4(4 +40y -10 y^2-4y^3+y^4)t^2
\right. \nonumber \\
&& \;\;\;\; \left.
+ \; 3y^3(4 + 13y - 4y^2)t^4 + 3y^2(5+2y)t^6 +(1+y)^2 t^8 \right\} \; ,
\end{eqnarray}
where $C_1 = 2 \alpha_s^2 |R'(0)|^2 /(3  \pi N_c m_c^5)$.
The color-octet piece $D^{(8)}(y,t)$ can be extracted from
Ref.\cite{bcfrags_pol} by taking the limit $r \to 0$,
\begin{equation}
\label{d8yt}
D^{(8)} (y,t) = 12 C_8 {(y-1)^2 t \over y^2 (t^2 + y^2)^4}
\Bigl[ 4y^2 + y(y+4)t^2 + t^4 \Bigr] \; ,
\end{equation}
with $C_8 = 3 \alpha_s^2 H_8^\prime /(256  N_c m_c)$
where $H_8^\prime$ is a nonperturbative parameter associated with the
color-octet mechanism for $P$-wave production \cite{bcfragp,bbl,gfragp}.
We will treat the
overall constants $C_1$ and $C_8$ as free parameters in our approach.
One notices that in terms of the variables $y$ and $t$, the leading order
results of the fragmentation functions $D_{T,L}(y,t)$ scale, {\it i.e.} they
do not depend on the heavy quark mass.

This motivates the introduction
of the generalized Falk-Peskin parameter
$w_{3/2}(y,t)$ that is a function of the two scaling variables $y$ and $t$.
The generalized polarized fragmentation functions $D(y,t)$
for the 8 helicity components of a $(1,2)$ doublet satisfy
a similar table like that given  in (\ref{pdoublet}) with
$w_{3/2}$ replaced by $w_{3/2}(y,t)$.
{}From the probability table for $P_1(h)$ in (\ref{pdoublet}), we see that
$D_T$ is proportional to $(1-w_{3/2})/8 \,+\,3w_{3/2}/8$, while
$D_L$ is proportional to $(1-w_{3/2})/4$. Thus $w_{3/2}(y,t)$ can be
defined in terms of the fragmentation functions for the spin-1 state by
\begin{equation}
\label{w32yt}
w_{3/2}(y,t) = {D_T(y,t) - {1 \over 2} D_L(y,t) \over
                D_T(y,t) + D_L(y,t)}  \; .
\end{equation}
Simple analytic expression for
$w_{3/2}(y,t)$ can be obtained from (\ref{dtyt})--(\ref{d8yt}) as
\begin{equation}
w_{3/2} (y,t)  = {n(y,t) + \epsilon \, h(y,t) \over
                  d(y,t) + 2 \epsilon \, h(y,t)}  \; \; ,
\end{equation}
with
$n(y,t) = 6 y^2 (y-1)^2 t^2 \Bigl[ 4y^2 + y(y+4)t^2 + t^4 \Bigr] $,
$d(y,t) = 2 (t^2 + y^2)  \Bigl[ y^4(y-4)^2 + y^3(24+y-4y^2+y^3)t^2
+ y^2 (17-2y+2y^2)t^4 + (1+y)^2 t^6 \Bigr]$,
$h(y,t) = 9 y^2 (t^2 + y^2)^2 \Bigl[ 4y^2 + y(y+4)t^2 + t^4  \Bigr]$,
and $\epsilon = C_8/C_1$.

We can define probabilities $w_{3/2}(t)$ and $w_{3/2}(y)$ that depend only
on a single scaling variable by an expression analogous to
(\ref{w32yt}), except with $D_{T,L}(y,t)$ replaced by $D_{T,L}(t)$ and
$D_{T,L}(y)$ respectively.
Integrating the fragmentation functions in (\ref{dtyt})--(\ref{d8yt})
over $y$ and forming a similar ratio like (\ref{w32yt}),
we obtain $w_{3/2}(t)$:
\begin{equation}
\label{w32t}
w_{3/2} (t) = {9 \over 80} \; \left(
{n_1 + n_2 \arctan (t) + n_3 \log (1+t^2)
\over d_1 + d_2 \arctan (t) + d_3 \log (1+t^2)} \right) \; ,
\end{equation}
with
$n_1 = t \Bigl[ 630 - 5 \, (521 - 240 \, \epsilon) \, t^2 + (231 - 2440 \,
\epsilon) \, t^4
\Bigr] \; ,
n_2 = - 5 \Bigl[ 126 - 15 \, (49 - 16 \, \epsilon) \, t^2 + 8 \,
(20 - 99 \, \epsilon) \, t^4
+ (5+72 \epsilon) \, t^6 \Bigr] \; ,
n_3 =  - 320 \, t \Bigl[ 4 - 2 \, ( 2 - 3 \, \epsilon) \, t^2 - 3 \,
\epsilon \, t^4 \Bigr]$,
and
$d_1 = t \Bigl[ 105 - 2 \, (406 - 135 \, \epsilon) \, t^2
+ (79 - 549 \, \epsilon) \, t^4
\Bigr] \; ,
d_2 = - 3 \Bigl[ 35 - 5 \, (71 - 18 \, \epsilon) \, t^2
+ 3 \, (31 - 99 \, \epsilon) \, t^4
+ 3 \, (1 + 9 \, \epsilon) \, t^6 \Bigr] \; ,
d_3 = - 72 \, t \Bigl[ 4 - 6 \, (1 - \epsilon) \, t^2 - 3 \, \epsilon \,
t^4 \Bigr]$.
The curve of $w_{3/2}(t)$ versus $t$ is plotted in
Fig.2 for the two cases of $\epsilon$ = 0  (color-singlet
dominance) and $\epsilon = \infty$ (color-octet dominance).

Similarly, by integrating the fragmentation functions in
(\ref{dtyt})--(\ref{d8yt}) over $t$ and forming the ratio
like (\ref{w32yt}), we obtain $w_{3/2}(y)$:
\begin{equation}
\label{w32y}
w_{3/2}(y) = {1 \over 10} \;
{(y-1)^2(12+8y+5y^2) + 15 \, \epsilon \, y^2(8+4y+3y^2) \over
(8+4y^2+y^4)+ 3 \, \epsilon \, y^2(8+4y+3y^2)} \; .
\end{equation}
The curve of $w_{3/2}(y)$ versus $y$ is plotted in Fig.3 for the
two cases of $\epsilon$ = 0 and $\infty$.

The original Falk-Peskin parameter $w_{3/2}$ is given by (\ref{w32yt}),
with the numerator and denominator integrated over both $y$ and $t$.
The total fragmentation probabilities are
$\int_1^\infty dy \int_0^\infty dt D_T(y,t) = (86/315)C_1 + (16/5) C_8$ and
$\int_1^\infty dy \int_0^\infty dt D_L(y,t) = (17/63)C_1 + (8/5) C_8$.
These imply
\begin{equation}
\label{w32}
w_{3/2} = {1 \over 6} \left(
{29 + 504 \, \epsilon \over 19 + 168 \, \epsilon} \right) \; .
\end{equation}
We note that setting $\epsilon = 0$ in (\ref{w32}) gives
$w_{3/2} \to 29/114$, a result first derived by Chen and Wise \cite{chenwise}.
While the Falk-Peskin parameter $w_{3/2}$ is independent of QCD evolution,
its generalizations $w_{3/2}(y,t)$, $w_{3/2}(t)$, and $w_{3/2}(y)$
given above are evaluated at a subtraction point around the heavy quark mass
scale.  Their values at a higher
scale can be determined by the usual Altarelli-Parisi evolution of the
fragmentation functions.

The Falk-Peskin parameter $w_{3/2}$ determines all the helicity probabilities
for the fragmentation of a heavy quark into a heavy meson whose light
degrees of freedom have angular momentum 3/2. In this Letter, we have pointed
out that $w_{3/2}$ can have nontrivial dependence on the scaling
variables $y$ and $t$ corresponding to the longitudinal
momentum fraction $z$ and the transverse momentum $p_\bot$ of the meson
relative to the heavy quark jet. We have calculated $w_{3/2}$ as a function
of $y$ and $t$ for the $m_b \to \infty$ limit of the perturbative QCD
fragmentation functions for the production of $P$-wave
$(b \bar c)$ mesons. We find that the probability $w_{3/2}$ has significant
dependence on $y$ and $t$, varying over the range  from 0 to 1/2.
The PQCD fragmentation functions can be used as a model for the fragmentation
of a heavy quark into heavy-light mesons, and applied to the fragmentation
processes $c \to D^{**}$ and $b \to B^{**}$.
The prediction $w_{3/2} \leq 1/2$ of this model implies that
light degrees of freedom with helicity $m_l = \pm 1/2$ always have
a larger population than the  maximum helicity states of $m_l = \pm 3/2$.
This prediction supports the speculation of Falk and Peskin \cite{falkpeskin}
that the angular momentum of the light degrees of freedom
with $j_l = 3/2$ prefers to align transverse to,
rather than along, the fragmentation axis.
This spin alignment of the heavy quark can be
detected by measuring the anisotropies of the decay products from
the hadronic transitions between the two doublets
$(1^+,2^+)$ and  $(0^-,1^-)$ \cite{falkpeskin}.
The PQCD fragmentation model predicts that
these anisotropies vary significantly with $y$ and $t$.
In the case of charm quark fragmentation into $D^{**}$,
an upper bound of $w_{3/2} \leq 0.25$ at the 90\%
confidence level has been deduced from the existing
experimental data \cite{falkpeskin}.
After integrating over $y$ and $t$, the prediction (\ref{w32}) of the
PQCD fragmentation model with $\epsilon = 0$ is
$w_{3/2} \approx 0.25$, which suggests that present experiments may be close
to observing a nonzero value for $w_{3/2}$.
String models of fragmentation tend to give significantly larger
values of $w_{3/2}$ and may already have been ruled out \cite{falkpeskin}.
We conclude that future measurements of the Falk-Peskin probability $w_{3/2}$
for the charm and bottom systems, including the dependence of $w_{3/2}$
on the scaling variables $y$ and $t$,  can provide valuable insights
into the dynamics of heavy quark fragmentation.

\medskip

I am grateful to Eric Braaten for valuable conversations and
Mark Wise for useful communication.
This work was supported in part by the U.S. Department of Energy,
Division of High Energy Physics under Grants DE-FG03-91ER40674.

\medskip


\medskip



\noindent{\bf Figure Captions}
\begin{enumerate}
\item Leading order diagram for the heavy $b$ quark to fragment into a
$P$-wave $(b \bar  c)$ bound state as a model for heavy quark
fragmentation into excited heavy-light meson.
The outgoing momenta are $(1-r)p + \delta$, $rp - \delta$, and
$p'$ for the $b$, $\bar c$, and $c$, respectively. $\delta$
is the relative momentum of the $b$ and $\bar c$.
\item $w_{3/2}(t)$ versus the rescaled transverse momentum
$t$ at the heavy quark mass scale.
\item $w_{3/2}(y)$ versus the scaling variable $y$ at
the heavy quark mass scale.
\end{enumerate}

\vfill\eject


\begin{thebibliography}{99}
%
\bibitem{wiseguys}
{N. Isgur and M. B. Wise, Phys. Lett. {\bf B232}, 113 (1989);
{\bf B237}, 527 (1990).}
%
\bibitem{hqet}
{E. Eichten and B. Hill, Phys. Lett. {\bf B234}, 511 (1990);
H. Georgi, Phys. Lett. {\bf B240}, 447 (1990).}
%
\bibitem{falkpeskin}
{A. F. Falk and M. E. Peskin, Phys. Rev. {\bf D50}, 3320 (1994).}
%
\bibitem{spectrum}
{N. Isgur  and M. B. Wise, Phys. Rev. Lett. {\bf 66}, 1130 (1991).}
%
\bibitem{swave}
{E. Braaten, K. Cheung, S. Fleming, and T. C. Yuan, in preparation.}
%
\bibitem{bcfrags}
{E. Braaten, K. Cheung, and T. C. Yuan, Phys. Rev. {\bf D48}, R5049 (1993).}
%
\bibitem{bcfragp}
{T. C. Yuan, U. C. Davis preprint, UCD-94-2, May (1994).}
%
\bibitem{chen}
{C.-H. Chang and Y.-Q. Chen, Phys. Lett. {\bf B284}, 127 (1992),
Phys. Rev. {\bf D46}, 3845 (1992);
Y.-Q. Chen, Phys. Rev. {\bf D48}, 5158 (1993).}
%
\bibitem{bbl}
{G. T. Bodwin, E. Braaten, and G. P. Lepage, Phys. Rev. {\bf D46},
R1914 (1992); Argonne preprint, ANL-HEP-94-24;
G. T. Bodwin, E. Braaten, T. C. Yuan, and G. P. Lepage,
Phys. Rev. {\bf D46}, R3703 (1992).}
%
\bibitem{gfragp}
{E. Braaten and T. C. Yuan, Fermilab preprint, FERMILAB-PUB-94/040-T (1994).}
%
\bibitem{jaffe}
{R. L. Jaffe and L. Randall, Nucl. Phys. {\bf B412}, 79 (1994).}
%
\bibitem{bcfrags_pol}
{K. Cheung and T. C. Yuan, U. C. Davis preprint, UCD-94-4, April (1994),
to appear in Phys. Rev. {\bf D}.}
%
\bibitem{chenwise}
{Y.-Q. Chen and M. B. Wise, Caltech preprint, CALT-68-1928, (1994).}
%
\end{thebibliography}
\end{document}